\documentclass[aip,prb,showpacs,twocolumn]{revtex4}

\usepackage{graphics,graphicx,amssymb,amsmath,color}
\begin{document}

\title{Synthesis and Characterization of Single Crystal Samples of Spin-$\frac{1}{2}$ Kagome Lattice Antiferromagnets in the Zn-Paratacamite Family Zn$_{x}$Cu$_{4-x}$(OH)$_{6}$Cl$_{2}$}
\author{T.H.~Han$^{1,\ddag}$}
\author{J.~S.~Helton$^{1,2}$}
\author{S.~Chu$^{3}$}
\author{A.~Prodi$^{1,\dagger}$}
\author{D.~K.~Singh$^{2,}$$^{4}$}
\author{C.~Mazzoli$^{5}$}
\author{P.~M\"{u}ller$^{6}$}
\author{D.~G.~Nocera$^{6}$}
\author{Y.~S.~Lee$^{1,\ddag}$}
\affiliation{$^{1}$Department of Physics, Massachusetts Institute
of Technology, Cambridge, Massachusetts 02139, USA}\affiliation{$^{2}$NIST Center for Neutron Research, Gaithersburg, Maryland 20899, USA}\affiliation{$^{3}$Center for Materials Science and Engineering, Massachusetts Institute of Technology, Cambridge, Massachusetts 02139, USA}\affiliation{$^{4}$Department of Materials Science and Engineering, University of Maryland, College Park, Maryland 20742, USA
}\affiliation{$^{5}$European Synchrotron Radiation Facility, 38043 Grenoble, France} \affiliation{$^{6}$Department of Chemistry, Massachusetts Institute of Technology, Cambridge, Massachusetts 02139, USA}
\date{\today}


\begin{abstract}
The Zn-paratacamite family, Zn$_{x}$Cu$_{4-x}$(OH)$_{6}$Cl$_{2}$ for
$x \, \geq$ 0.33, is an ideal system for studying spin-$\frac{1}{2}$
frustrated magnetism in the form of antiferromagnetic Cu$^{2+}$
kagome planes. Here we report a new synthesis method by which high
quality millimeter-sized single crystals of Zn-paratacamite have
been produced.  These crystals have been characterized by metal
analysis, x-ray diffraction, neutron diffraction, and thermodynamic
measurements. The $x$~=~1 member of the series displays a magnetic
susceptibility that is slightly anisotropic at high temperatures
with $\chi_{c} \, > \, \chi_{ab}$. Neutron and synchrotron x-ray diffraction
experiments confirm the quality of these $x$~=~1 single crystals and
indicate no obvious structural transition down to temperatures of
$T=2$~K.
\end{abstract}

 \maketitle

Geometrically frustrated magnetism\cite{Ramirez,MisguichLhuillier}
is a forefront area of research in condensed matter physics, as such
systems offer a unique terrain in which to search for novel magnetic
ground states.  The spin-$\frac{1}{2}$ nearest-neighbor Heisenberg
antiferromagnet on the kagome lattice, which consists of corner
sharing triangles, is a particulary promising system in which to
search for unique quantum phases including the ``resonating valence
bond'' (RVB) state proposed by Anderson\cite{Anderson1973} or other
quantum spin liquid states.  A broad theoretical and numerical
consensus has emerged that the ground state of this system is not
magnetically ordered\cite{Zeng,Marston,SinghHuse1992,Sachdev}, with
a variety of proposed ground states including gapped spin
liquids\cite{Waldtmann}, gapless spin liquids\cite{Ran,Ryu}, and
valence bond solid (VBS) states\cite{SinghHuse2007,Nikolic}.
However, experimental investigation on this system has long been
hampered by the fact that most early realizations of the kagome
lattice antiferromagnet feature either large spins or structural
distortions.

The material ZnCu$_{3}$(OH)$_{6}$Cl$_{2}$\cite{Braithwaite,Shores}  is among the best realizations of a
spin-$\frac{1}{2}$ kagome lattice antiferromagnet yet synthesized.
This material is a member of the Zn-paratacamite family
Zn$_{x}$Cu$_{4-x}$(OH)$_{6}$Cl$_{2}$ with $x=1$.  With lattice
parameters $a$ = $b$ = 6.83 {\AA} and $c$ = 14.05 {\AA},
ZnCu$_{3}$(OH)$_{6}$Cl$_{2}$ is rhombohedral (trigonal setting) and consists of
kagome lattice planes of spin-$\frac{1}{2}$ Cu$^{2+}$ ions separated
by layers of non-magnetic Zn$^{2+}$ ions as shown in Fig.1(a).
Measurements on powder samples\cite{MendelsJPSJ}
have found no sign of long range order or spin freezing down to
temperatures of 50~mK\cite{Helton,Mendels,Ofer}, despite a strong
antiferromagnetic superexchange interaction of $J \, \approx$~17~meV
and a Curie-Weiss temperature of $\Theta_{CW} \, = \, -300 \, \pm$
20 K. There is no evidence of a spin gap down to at least
J/200\cite{Helton,Olariu,Wulferding}. The most significant
deviations of this material from the idealized model are likely the
presence of about 5\% weakly-coupled Cu$^{2+}$ ions lying on out-of-plane metallic
sites, which may be responsible for the Curie-like susceptibility at
low temperatures, and Dzyaloshinksii-Moriya or exchange anisotropy interactions.  Recent anomalous x-ray diffraction measurements indicate that dilution of the kagome plane sites with Zn ions is not significant\cite{Freedman}.  Studies on Zn-paratacamite
samples\cite{Mendels,SHLee,deVries} with $x \, < \, 1$ can be useful
in determining the effect of any out-of-plane impurities that might
be present even in the nominal $x = 1$ samples, as well as serving as model
systems for weakly coupled kagome planes.

\begin{figure}
\centering
\includegraphics[width=8.7cm]{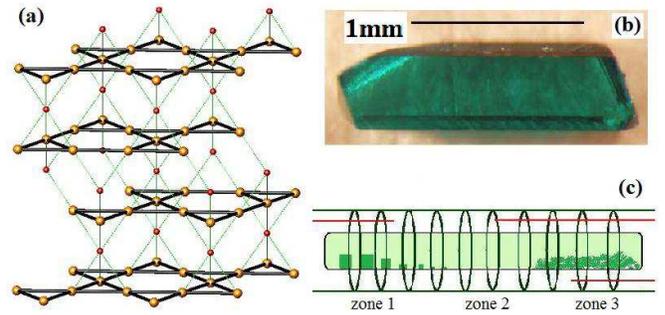} \vspace{-4mm}
\caption{(color online) (a) Structure of ZnCu$_{3}$(OH)$_{6}$Cl$_{2}$ with only
Cu$^{2+}$(large brown spheres) and Zn$^{2+}$(small red spheres)
displayed. The Cu-Cu bonds (thick black solid lines) are all
equivalent as are the Cu-Zn bonds (thin green dotted lines). (b) A
single crystal sample of ZnCu$_{3}$(OH)$_{6}$Cl$_{2}$. (c) A schematic of three
zone furnaces. The red bars indicate the positions of the
thermocouples.} \vspace{-4mm} \label{Figure1}
\end{figure}

Previous studies of Zn-paratacamite have utilized powder samples
grown by hydrothermal methods.  The failure of these methods to
produce large single crystals has been ascribed partly to the low
decomposition temperature of ZnCu$_{3}$(OH)$_{6}$Cl$_{2}$. However, the primary
reason for the lack of single crystal synthesis is likely the fact
that the reported synthesis\cite{Shores} produces bubbles of CO$_{2}$, resulting in an unstable
crystallization environment.  Further understanding of the spin
behavior of ZnCu$_{3}$(OH)$_{6}$Cl$_{2}$, and further insight into the ground
state of the spin-$\frac{1}{2}$ kagome lattice antiferromagnet, will
require studies on single crystal samples.  Here we report a new
synthesis method by which high quality millimeter-sized single
crystals have been successfully produced.  These samples have been
characterized by a variety of measurements.

Single crystal samples of Zn-paratacamite,
Zn$_{x}$Cu$_{4-x}$(OH)$_{6}$Cl$_{2}$, were grown hydrothermally in
furnaces that were setup similarly to those used to grow small
single crystal samples of the atacamite family\cite{Chu_unpub} and Mg$_{x}$Cu$_{4-x}$(OH)$_{6}$Cl$_{2}$\cite{Chu}.  Here, starting
materials of CuO, ZnCl$_{2}$, and H$_{2}$O, in amounts listed in
Table~\ref{SynthesisTable}, were charged into a fused quartz tube
(ID
 6 mm, OD 13 mm for x=0.8 and 1.0 or ID
 9 mm, OD 15 mm for x=0.9).  The quartz tube was sealed after purging air with a mechanical pump.  The sealed quartz tube was prereacted for two days in a box furnace at 185 $^{\circ}$C.  After prereaction, a green-blue microcrystalline powder was formed.  Powder x-ray diffraction measurements of this product have indicated the presence of Zn$_{x}$Cu$_{4-x}$(OH)$_{6}$Cl$_{2}$.  This shows successful synthesis by the reaction
\begin{multline}
(4-x)\mbox{CuO} \, + \, \mbox{ZnCl}_{2} \, + \, 3\mbox{H}_{2}\mbox{O} \, \rightarrow \, \\
\mbox{Zn}_{x}\mbox{Cu}_{4-x}(\mbox{OH})_{6}\mbox{Cl}_{2} \, + \, (1-x)\mbox{ZnO}.
\end{multline}
This synthesis without the production of CO$_{2}$ suggests the
possibility of an environment stable enough for single crystal
growth.

Millimeter-sized single crystals were synthesized through a
recrystallization process in a three-zone gradient tube furnace.  A
schematic of such a furnace is shown in Fig.~\ref{Figure1}(c).  The
sealed, prereacted quartz tubes were placed horizontally into the
furnace at room temperature.  The furnace temperature was
isotropically increased to a fixed temperature, ranging from
165~$^{\circ}$C to 180~$^{\circ}$C in various reactions.  The
temperature of the cold end was then slowly lowered. The sample and
all growth parameters were undisturbed for roughly 20 weeks until
large crystals were formed at the cold end.  In the region where the
crystals nucleated and grew, the temperature gradient was measured
to be approximately 1~$^{\circ}$C/cm.  At the end of the synthesis,
the sample tubes were cooled down to room temperature at
1~$^{\circ}$C/min. Crystals were then rinsed with deionized water,
dried in air and kept in a desiccator for storage. No decomposition
of the crystals has been observed in air, water, or acetone. Precise
control of the starting concentrations of CuO and ZnCl$_{2}$ allows
for synthesis of samples with variable Zn concentration, $x$. Data
on crystals with $x$ = 0.8, 0.9, and 1.0 are shown in
Table~\ref{SynthesisTable}.  More syntheses than listed in
Table~\ref{SynthesisTable} were performed with ZnCl$_{2}$ to CuO
molar ratios ranging from 2 to 10 and with ZnCl$_{2}$ to H$_{2}$O
concentrations ranging from 1.2~mmol/ml to~7.7 mmol/ml. However,
the $x$ values of the final products were fairly stable over this
range of starting concentrations.  At a fixed  ZnCl$_{2}$ to CuO
ratio, the $x$ value of the product increased with increasing
ZnCl$_{2}$ concentration, from an $x$ value of 0.8 with a ZnCl$_{2}$
concentration of 1.2 mmol/ml to an $x$ value of 1.0 with a
ZnCl$_{2}$ concentration of 2.8 mmol/ml.  Products with an $x$
value of 1.0 were also obtained for starting ZnCl$_{2}$
concentrations up to 5.6~mmol/ml, while even higher ZnCl$_{2}$
concentrations resulted in a slight lowering of $x$.  This
diminishment of $x$ with very high ZnCl$_{2}$ concentrations is
likely due to a more acidic pH in those reactions which dissolves
more CuO. The ZnCl$_{2}$ to CuO molar ratio of the starting products
had no obvious effect on the $x$ values of the product over the
range of syntheses performed.  The compositions of the crystals were
measured by metal analysis taken with an inductively coupled plasma
atomic emission spectrometer (ICP-AES) with an error of $\pm$0.04 on
x. Standards were prepared from commercially purchased solutions
from Sigma-Aldrich, specific for ICP-AES measurements and designated
as Trace SELECT grade or better. Five to ten well rinsed small
single crystals from each synthesis tube, approximately 0.1~mg each,
were dissolved into 2\%~w/w nitric acid for measurement. Unlike
powder samples, the ease to rinse single crystals dramatically
reduced the ambiguity from possible chemical contamination. The
relative amounts of Cu and Zn determined from ICP metal analysis
were used to calculate the values of $x$ listed in
Table~\ref{SynthesisTable}.  Our nominal $x=1$ sample has previously been determined to have structural composition (Zn$_{0.85}$Cu$_{0.15}$)Cu$_3$(OH)$_6$Cl$_2$ via anomalous x-ray scattering.\cite{Freedman}  Here, the kagome planes are fully occupied with Cu, and anti-site disorder\cite{Singh10,Devries09} with Zn on the Cu kagome site is not apparent.

\begin{table}
\caption{Growth and crystallography data. All samples have
rhombohedral crystal system in R$\bar{3}$m space group ($\alpha$=$\beta$=90$^{\circ}$, $\gamma$=120$^{\circ}$. Single crystal
x-ray diffraction was performed at $T$~=~100 K and refined by
full-matrix least-squares on F$^2$ with goodness-of-fit (GOF) listed.
Curie-Weiss temperatures were determined from high temperature
susceptibilities.}
\begin{tabular}{| c | c | c | c |}
\hline \hline
 & $x \, = \, 0.8$ & $x \, = \, 0.9$ & $x \, = \, 1.0$ \\
\hline
Starting& CuO (0.130) & CuO (0.346) & CuO (0.235)\\
materials (g)& ZnCl$_{2}$ (0.686) & ZnCl$_{2}$ (2.985) & ZnCl$_{2}$ (2.015)\\
& H$_{2}$O (4.0 ml) & H$_{2}$O (10.0 ml) & H$_{2}$O (4.5 ml)\\
\hline
Hot zone temp. & 165 $^{\circ}$C & 165 $^{\circ}$C & 180 $^{\circ}$C \\
\hline
$a$ & 6.8300(13) {\AA} & 6.8345(9) {\AA} & 6.8332(12) {\AA} \\ $b$ & 6.8300(13) {\AA} & 6.8345(9) {\AA} & 6.8332(12) {\AA} \\
$c$ & 14.029(3) {\AA} & 14.0538(19) {\AA} & 14.066(2) {\AA} \\
Volume ({\AA}$^{3})$ & 566.77(19) & 568.51(13) & 568.80(17)\\
Density, $\rho$ & 3.765 g/cm$^{3}$ & 3.759 g/cm$^{3}$ & 3.755 g/cm$^{3}$\\
total reflections  & 2673 & 3718 & 5908\\
indep reflections  & 238 & 225 & 240\\
GOF  & 1.275 & 1.264 & 1.237\\
\hline
$\Theta_{CW}$ & -266(10) K & -290(10) K & -296(10) K \\

\hline \hline
\end{tabular}
\label{SynthesisTable}
\end{table}

Single crystal x-ray diffraction was performed on a three-circle
diffractometer coupled to a CCD detector.  All samples were refined
in the rhombohedral space group R$\bar{3}$m (trigonal setting) and
with lattice constants consistent with previous reports.  The
largest (1 to 10 mm$^{3}$) crystals typically grow as a bar-shape similar to the crystal shown in Fig.~1(b).  From x-ray diffraction on more than 30 crystals, all of
the side long faces were indexed as (1~0~1) (the faces were normal
to the (1~0~1) reciprocal lattice vector).  Some of the smaller ($<$
0.5 mm$^{3}$) crystals were octahedrally shaped, with all eight
faces indexed as (1~0~1).  Based on these observations, we propose
the following growth process: during the early stage of
crystallization, primitive micrometer-sized crystals form as
twelve-faced polyhedra due to the symmetry of the (1~0~1)
directions.  As the crystals grow larger in size, eight of the
twelve (1~0~1) faces, possibly due to the specific local
hydrodynamic condition, grow faster which transforms the crystals
into larger, sub-millimeter octahedra. Eventually, four of these
eight (1~0~1) faces form the four large side faces of
millimeter-sized bar-shaped crystals.

\begin{figure}
\centering
\includegraphics[width=8.8cm]{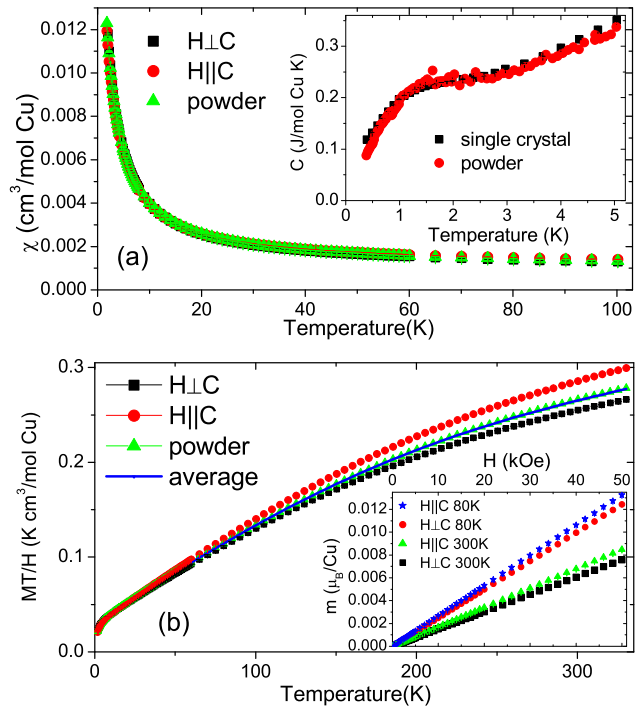} \vspace{-4mm}
\caption{(color online) (a) The bulk susceptibility, $M/H$, of
an $x=1$ single crystal as a function of temperature.  Also
plotted are results from powder samples. Inset: Low temperature
specific heat of an $x=1$ single crystal compared
with powder data (with $H=0$). (b) Plots of $MT/H$ versus
temperature, showing the anisotropy $\chi_{c} \, > \, \chi_{ab}$ at
high temperatures. Inset: M versus H curves at 80~K and 300~K along
different directions.} \vspace{-4mm} \label{Susceptibility}
\end{figure}

The single crystal samples obtained by this synthesis were
characterized by a variety of methods. Single crystal susceptibility
measurements of an $x=1$ sample were performed on a SQUID
magnetometer (Quantum Design) using a 55.5~mg sample (different from the one in Fig.~\ref{Figure1}(b)) with an almost cubic shape (2.3 mm $\times$ 2.5 mm $\times$ 2.7 mm). Figure.~\ref{Susceptibility}(a) shows
the bulk susceptibility, $\chi \equiv M/H$, as a function of
temperature. The susceptibility of the crystal closely follows that
measured on a powder sample with $x=1$ (also plotted), which
confirms that the stoichiometry and homogeneity of the single
crystals match that of previously measured powders. The inset of
panel (a) shows the specific heat of an $x=1$ single crystal measured at $\mu_{0}H$~=~0~T and compared with data from a powder sample.  Again, the two curves show good agreement, where the higher quality of single crystal measurements is likely due to the better thermal contact with the sample holder. In
Fig.~\ref{Susceptibility}(b), we show the magnetic susceptibility
plotted as $M T/H$, with an applied magnetic field of
$\mu_{0}H$~=~1~T aligned along the different crystallographic axes,
($\chi_{c}$) and($\chi_{ab}$).  Powder data are also plotted
($\chi_{powder}$) for comparison. The calculated powder averaged
value of the single crystal data ($\chi_{av} \, \equiv \,
\frac{2}{3}\chi_{ab} + \frac{1}{3}\chi_{c}$) is indicated by the
solid line.  As expected, the calculated $\chi_{av}$ and
$\chi_{powder}$ show good agreement.  At high temperatures, the
single crystal data show clear magnetic anisotropy with $\chi_{c} \,
> \, \chi_{ab}$. The inset to panel (b) shows the high temperature
(80~K and 300~K) magnetization; the anisotropy is field independent,
with $M_{c}/M_{ab}$ a constant up to $\mu_{0}H$~=~5~T.  This high
temperature anisotropy is consistent with the qualitative
results obtained from powder samples that were partially
oriented in a magnetic field\cite{Ofer2} and $\mu$SR measurement on single crystal samples\cite{Ofer3}. The anisotropy of the magnetization
should play a role in elucidating the effects that the
Dzyaloshinskii-Moriya interaction, exchange anisotropy, or out-of-plane impurities have on the low temperature physics of the material\cite{Rigol1,Rigol2}. The bulk susceptibility of crystals with $x=0.8$ and $0.9$ are also consistent with powder samples (not shown). A detailed study of the susceptibility and specific heat of samples with different $x$-values is a topic of ongoing investigation.

Among the various predicted ground states for the spin-$\frac{1}{2}$
nearest-neightbor Heisenberg kagome antiferromagnet, several
theoretical studies have suggested a 36-site valence bond
solid\cite{Nikolic,SinghHuse2007}. This proposed ground state
features a $\sqrt{12}\times\sqrt{12}$ enlargement of the unit cell,
with the 36 spin sites paired into 18 nearest-neighbor dimer
singlets, where 6 dimers lie around a central pinwheel configuration while another 6 lie around hexagons.  The two possible coverings of the pinwheel are degenerate to high order, while sets of three dimers can resonate around a perfect
hexagon\cite{SinghHuse2007}.  The other 6 dimers are presumed to be
static.  It has been suggested\cite{Lawler} that a VBS order might
lead to a slight structural distortion in which the distance between
two magnetic ions paired in a singlet is diminished.  Such a
distortion, if large enough, could in principle be measured in a
synchrotron x-ray experiment.

\begin{figure}
\centering
\includegraphics[width=8.9cm]{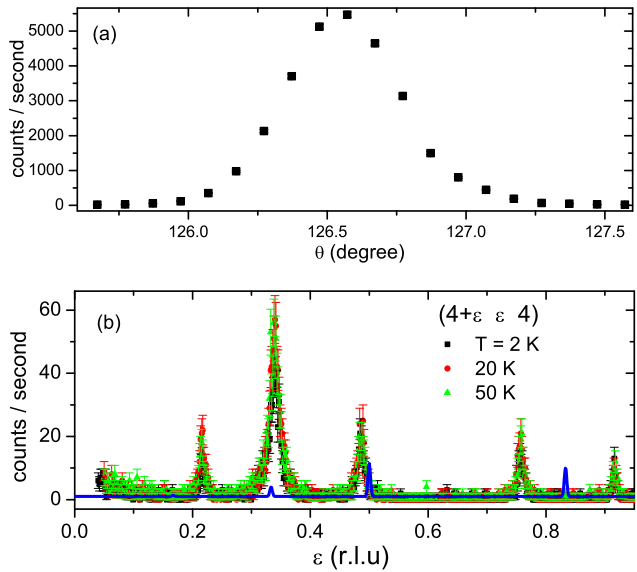} \vspace{-4mm}
\caption{(color online) (a) Neutron diffraction $\theta$-scan
through the (1~1~0) Bragg reflection of an $x=1$ single
crystal measured with the SPINS neutron spectrometer. (b)
Synchrotron x-ray diffraction intensity of a scan along the
(4+$\epsilon$~$\epsilon$~4) direction at three temperatures, 2~K,
20~K and 50~K.  The extremely weak peaks observed in this range
arise from powder contamination and have no temperature dependence.
The lines are a simulation of the superlattice peaks arising from a
36-site valence bond solid with a 1\% reduction in the bond length
of the static dimer pairs, as discussed in the text.} \vspace{-4mm}
\label{ESRF}
\end{figure}

We performed diffraction measurements on our single crystals with
$x=1$.  Figure.~\ref{ESRF}(a) shows a $\theta$-scan through the (1~1~0)
Bragg reflection measured with neutron diffraction using the SPINS
spectrometer at the NIST Center for Neutron Research.  The width of
the scan of 0.5$^{\circ}$, which is resolution-limited, attests to
the crystal homogeneity.  A high-resolution x-ray diffraction
experiment on a small $x~=~1.0$ single crystal was performed on the
ID20 beamline of the European Synchrotron Radiation Facility (ESRF).
The scattering was performed in reflection geometry with the (1~0~1)
reflection roughly perpendicular to the mount, with x-rays of energy
of 8.979 keV ($\lambda$~=~1.381 {\AA}). The sample was cooled with a
closed-cycle displex and was mounted on a four-circle goniometer.  The measurements indicate that the sample remains in space group R$\bar{3}$m down to the lowest
measured temperature $T\simeq2$~K.  The high x-ray flux available at
a synchrotron is ideal to look for very subtle superlattice
reflections that would arise if a VBS ground state resulted in a
structural distortion.  The 36-site VBS would lead to an enlarged unit
cell that is a factor of $\sqrt{12}$ longer on each side and rotated
90$^{\circ}$ from the original unit cell.  We searched for the
superlattice reflections along the high-symmetry (1~1~0) direction
between the (4~0~4) and (5~1~4) peaks. This scan, shown in
Fig.~\ref{ESRF}(b), showed no observable superlattice peaks above the
background.  The only scattering features in this range were very
weak powder peaks (roughly 250,000 times weaker than the strongest
lattice reflections). These peaks were confirmed to arise from
powder through $\theta$-scans; they also displayed no temperature
dependence and were somewhat broader than resolution. Contamination
from these powder peaks, although exceptionally weak, is likely the
limiting factor in setting an upper bound on the possibility of any
superlattice reflections.  The blue lines in Fig.~\ref{ESRF}(b) are
a calculation of the superlattice peaks expected if the copper ions
making up the 6 static dimer pairs per supercell (dimers that are in
neither a pinwheel nor a perfect hexagon) were to move toward one
another such that the Cu-Cu distance was reduced by 1\%. It was
assumed that the centering of the enlarged unit cells on adjacent
kagome planes is random, so that the superlattice peaks will
actually be rods of scattering parallel to $c$. The lack of observed
superlattice peaks at the expected positions indicates that any
structural distortions due to this proposed supercell are below the
1\% level. Note that the intensity of the superlattice peaks is
proportional to the displacement squared in the limit of small
displacement. Additional mesh scans (not shown) found no evidence of
other superlattice peaks, as might arise from VBS states with
different enlarged unit cells\cite{Lawler}.  Of course, this
experiment cannot rule out the presence of a VBS ground state that
results in very little or no structural change.

In summary, high quality single crystals of Zn$_{x}$Cu$_{4-x}$(OH)$_{6}$Cl$_{2}$ have been synthesized and characterized. The bulk properties of $x=1$ single crystals are consistent with the previously published powder results.  The susceptibility measured along different crystallographic directions shows clear anisotropy. This indicates the presence of additional terms in the spin Hamiltonian, such as a small Dzyaloshinskii-Moriya interaction or exchange anisotropy. Synchrotron x-ray scattering experiments did not show evidence for the emergence of superlattice peaks at low temperatures.  Hence, any possible lattice distortions associated with a valence bond solid are subtle, if they exist. Clearly, further measurements on these single crystal samples should help reveal the physics of the $S=1/2$ kagome ground-state.

We thank E.A. Nytko, D. Freedman, and T. McQueen for useful
discussions. The work at MIT was supported by the Department of
Energy (DOE) under Grant No. DE-FG02-07ER46134.  This work utilized
facilities supported in part by the National Science Foundation
under Agreement No. DMR-0454672.

$^{\dagger}$ Present address: Niels Bohr Institute, University of Copenhagen, Denmark\\
$^{\ddag}$ email: tianheng@alum.mit.edu, younglee@mit.edu
\bibliography{Synthesis}
\end{document}